\documentclass{svproc}
\usepackage{url}

\usepackage{hyperref}
\hypersetup{
    colorlinks=true,
    linkcolor=blue,
    filecolor=magenta,      
    urlcolor=cyan,
    citecolor = red,
}

\usepackage{subfigure}

\usepackage{amsmath}
\usepackage{amssymb}

\usepackage{mathrsfs}

\usepackage{amsfonts}

\usepackage{epsfig}

\usepackage{bm}

\usepackage{mathrsfs}

\newcommand{\calD}{\mathcal{D}}

\newcommand{\cI}{\mathcal{I}}

\newcommand{\cO}{\mathcal{O}}

\newcommand{\bbE}{\mathbb{E}}

\newcommand{\bbR}{\mathbb{R}}

\newcommand{\bbZ}{\mathbb{Z}}

\begin{document}
\mainmatter              
\title{Randomized multilevel Monte Carlo \\
for embarrassingly parallel inference}
\titlerunning{rMLMC for inference}  
%
\author{Ajay Jasra\inst{1}\and 
Kody J.~H. Law\inst{2} \and 
Alexander Tarakanov\inst{2} \and
Fangyuan Yu\inst{1}}
\authorrunning{Jasra, Law, 
Yu} 
%
\tocauthor{Kody J.~H. Law}
\institute{
Computer, Electrical and Mathematical Sciences and Engineering Division, King Abdullah University of Science and Technology, Thuwal, 23955, KSA.\\
\email{ajay.jasra@kaust.edu.sa, fangyuan.yu@kaust.edu.sa}
\and 
Department of Mathematics,
University of Manchester, Manchester, M13 9PL, UK.\\
\email{kodylaw@gmail.com, tarakanov517@gmail.com }
}

\maketitle              

\begin{abstract}
This position paper summarizes a recently developed research program 
focused on inference in the context of data centric science and engineering
applications, and forecasts its trajectory forward over the next decade.
Often one endeavours in this context to learn complex systems in 
order to make more informed predictions and high stakes decisions
under uncertainty.
Some key challenges which must be met in this context are 
robustness, generalizability, and interpretability.
The Bayesian framework addresses these three challenges,
while bringing with it a fourth, undesirable feature: it is typically far 
more expensive than its deterministic counterparts.
In the 21st century, and increasingly over the past decade, 
a growing number of methods have emerged which allow one to leverage cheap 
low-fidelity models in order to precondition algorithms for 
performing inference with more expensive models 
and make Bayesian inference tractable in the context of 
high-dimensional and expensive models.
Notable examples are multilevel Monte Carlo (MLMC), 
multi-index Monte Carlo (MIMC), and their randomized counterparts (rMLMC),
which are able to provably achieve a dimension-independent (including $\infty-$dimension)
canonical complexity rate with respect to mean squared error (MSE) of $1/$MSE.
Some parallelizability is typically lost in an inference context, but 
recently this has been largely recovered via novel {\em double randomization} approaches.
Such an approach delivers independent and identically distributed 
samples of quantities of interest which are unbiased with respect to the 
{\em infinite resolution target distribution}.
Over the coming decade, this family of algorithms has the potential to transform 
data centric science and engineering, as well as classical machine learning applications
such as deep learning, by scaling up and scaling out fully Bayesian inference.
\keywords{Randomization Methods; Markov chain Monte Carlo; Bayesian Inference}
\end{abstract}

\section{Introduction}\label{sec:intro}

The Bayesian framework begins with a statistical model characterizing the 
causal relationship between various variables, parameters, and observations.
A canonical example in the context of inverse problems is
$$
y \sim N(G_\theta(u), \Gamma_\theta) \, , \quad u \sim N(m_\theta,C_\theta) \, , 
\quad \theta \sim \pi_0 \, ,
$$
where $N(m,C)$ denotes a Gaussian random variable with mean $m$ 
and covariance $C$, 
$G_\theta: U \rightarrow \bbR^m$ 
is the (typically nonlinear) parameter-to-observation map, 
$\theta \in \bbR^p$ is a vector of parameters with $\pi_0$ some distribution,
and the data is given in the form of {\em observations} $y$ \cite{stuart,tarantola}. 
Nothing precludes the case where $U$ is a function space, e.g. leading to a Gaussian process
prior above, but to avoid unnecessary technicalities, assume $U=\bbR^d$.
The objective is to {\em condition} the prior knowledge about $(u,\theta)$
with the observed data $y$ and recover a {\em posterior distribution}
$$
p(u, \theta | y) = \frac{p(u,\theta,y)}{p(y)} = 
\frac{p(y | u, \theta) p(u | \theta) p(\theta)}{\int_{U \times \bbR^p} p(y | u, \theta) p(u | \theta) p(\theta)dud\theta} \, .
$$
Often in the context above one may settle for a slightly simpler goal 
of identifying a point estimate $\theta^*$, e.g. $\theta^* =$ argmax$_\theta p(\theta | y)$ 
(which we note may require an intractable integration over $U$)
and targeting $p(u | y, \theta^*)$ instead.

In the context described above, one often only has access to an {\em approximation} 
of the map $G_\theta$, and potentially an approximation of the domain $U$,
which may in principle be infinite dimensional. 
One example is the numerical solution of a system of differential equations.
Other notable examples include surrogate models arising from reduced-physics or 
machine-learning-type approximations \cite{multifidelity}
or deep feedforward neural networks \cite{neal}. 
For the sake of concreteness the reader can keep this model in mind, 
however it is noted that the framework is much more general, for example the parameters $\theta$ 
can encode the causal relationship between latent variables via a graphical model such as a deep 
belief network or deep Boltzmann machine \cite{bishop,murphy}.

A concise statement of the general problem of Bayesian inference 
is that it requires 
exploration of a posterior distribution $\Pi$ 
from which one cannot obtain independent and identically distributed (i.i.d.) samples.
Specifically, the aim is  to compute 
quantities such as 
\begin{equation}\label{eq:bip}
\Pi_\theta(\varphi) := \int_U \varphi(u) \Pi_\theta(du) \, , \quad \varphi: U \rightarrow \bbR \, ,  
\end{equation}
where $\Pi_\theta(du) = \pi_\theta(u)\nu_\theta(du)$,  
$\nu_\theta(du)$ is either Lebesgue measure $\nu_\theta(du)=du$, or one can simulate from it,
$\pi_\theta(u) = \gamma_\theta(u)/\nu_\theta(\gamma_\theta)$, 
and given $u$ one can evaluate 
$\gamma_\theta(u)$ (or at least a non-negative unbiased estimator).
Markov chain Monte Carlo (MCMC)
and sequential Monte Carlo (SMC) samplers can be used for 
this \cite{robert2011short}. 
Considering the example above with $U=\bbR^d$, 
we may take $\nu(du)=du$ and then 
\begin{equation}\label{eq:bippart}
\gamma_\theta(u) = 
|\Gamma_\theta|^{-1/2}  |C_\theta|^{-1/2} 
\exp(-\frac12|\Gamma_\theta^{-1/2}(y-G_\theta(u))|^2 -\frac12|C_\theta^{-1/2}(u-m_\theta)|^2) \, ,
\end{equation}
where $|A|$ denotes the determinant for a matrix $A\in\bbR^n$.
Note we have used a subscript for $\theta$, as is typical in the statistics literature
to denote that everything is conditional on $\theta$, and note that the $\theta-$dependent 
constants are not necessary here, per se, but it is customary to define the un-normalized
target as the joint on $(u,y)$, such that $Z_\theta := \nu_\theta(\gamma_\theta) = p(y|\theta)$.
Also note that in \eqref{eq:bippart}, $u$ would be referred to as a {\em latent variable} in the statistics 
and machine learning literature, and so this setup corresponds to a complex 
{\em physics-informed (via $G_\theta$) unsupervised learning model}.
Labelled data problems like regression and classification \cite{nealregression}, 
as well as semi-supervised learning \cite{lawrence2004semi,zhu2003semi}, 
can also be naturally cast in a Bayesian framework.
In fact, if $G_\theta(u)$ is point-wise evaluation of
$u$, i.e. $G_\theta^i(u) = u(x^i)$, for {\em inputs} or {\em covariates} $x^i$ associated to labels $y^i$,
and one allows $U$ to be an infinite-dimensional (reproducing kernel)
Hilbert space, then standard Gaussian process (GP) regression has this form.
In infinite-dimensions there is no Lebesgue density, so 
\eqref{eq:bippart} does not make sense, but the marginal likelihood 
and posterior can both be computed in closed form thanks to the properties of GP
\cite{rasmussen2006gaussian}.
Alternatively, if $u$ are the parameters of a deep feedforward neural network \cite{neal} 
$f_\theta(\cdot; u)$, and $G_\theta^i(u)= f_\theta(x^i; u)$ 
with Gaussian prior on $u$, then one has a standard Bayesian neural network model \cite{neal,bishop}.

\subsection{The sweet and the bitter of Bayes}

Three challenges which are elegantly handled in a Bayesian framework 
are (a) robustness, (b) generalizability, and (c) interpretability \cite{sciml,ai4sci}.
Uncertainty quantification (UQ) has been a topic of great interest in science and engineering 
applications over the past decades, due to its ability to provide a more robust model \cite{uqhandbook,sciml}. 
A model which can extrapolate outside training data coverage is 
referred to as generalizable. Notice that via prior knowledge \eqref{eq:bip}
and the physical model, \eqref{eq:bippart} has this integrated capability by design.
Interpretability is the most heavily loaded word among the three desiderata.
Our definition is that the model (i) can be easily understood by the user \cite{xaiuk},
(ii) incorporates all data and domain knowledge available in a principled way \cite{xaiuk,aiuk},
and (iii) enables inference of causal relationships between latent and observed variables \cite{pearl}.
The natural question is then, 
``Why in the age of data doesn't everybody adopt Bayesian inference for all their learning requirements?''

The major hurdle to widespread adoption of a fully Bayesian treatment of learning
is the computational cost. 
Except for very special cases, such as GP
regression \cite{rasmussen2006gaussian}, the solution cannot be obtained in closed form. 
Point estimates, Laplace approximations \cite{rue2009approximate}, 
and variational methods \cite{jordan1999introduction,blei}
have therefore taken center stage, as they can yield acceptable results very quickly in many cases. 
In particular, for a strongly convex objective function, gradient descent 
achieves exponential convergence to a local minimizer, i.e. MSE $\propto \exp(-N)$ in $N$ steps.
Such point estimates are still suboptimal from a Bayesian perspective, 
as they lack UQ.
In terms of computation of \eqref{eq:bip}, 
Monte Carlo (MC) methods are able to achieve exact inference in \eqref{eq:bip} in general 
\cite{metropolis1949monte,robert2011short}.
In the case of i.i.d. sampling, MC methods achieve the canonical, dimension-independent
convergence rate of MSE $\propto 1/N$, for $N-$sample approximations, without any smoothness assumptions and
out-of-the-box\footnote{This is the same rate achieved by gradient descent for general non-convex smooth objective functions.
In fact, the success of deep neural networks for learning high-dimensional functions has been
attributed to this dimension-independence in \cite{weinan2020integrating}.}.
Quadrature methods \cite{quad} and quasi-MC \cite{quasi} are able to achieve improvements 
over MC rates, however the rates depend on the dimension and the smoothness of the integrand.

A curse of dimensionality can still hamper application of MC methods 
through the constant and the cost of simulation, meaning
it is rare to achieve canonical {\em complexity} of cost $\propto 1/$MSE for non-trivial applications.
Usually this is manifested in the form of a penalty in the exponent, so that cost 
$\propto$ MSE$^{-a}$, for $a>2$.
A notable exception is MLMC \cite{heinrich2001multilevel,giles2015multilevel} 
and MIMC \cite{haji2016multi} methods, and their randomized counterparts rMLMC 
\cite{rhee2015unbiased,vihola2018unbiased}
and rMIMC \cite{crisan2018unbiased}, which are able to achieve dimension-independent 
canonical complexity for a range of applications. 
These estimators are constructed by using 
a natural telescopic sum identity and constructing coupled increment estimators of decreasing variance. 
As an added bonus, the randomized versions eliminate discretization bias {\em entirely},
and deliver estimates with respect to the limiting {\em infinite-resolution distribution}.

In the context of inference problems, i.i.d. sampling is typically not possible and one 
must resort to MCMC or SMC \cite{robert2011short}. 
This makes application of (r)MLMC and (r)MIMC more complex. 
Over the past decade, there has been an explosion of interest in applying
these methods to inference, e.g. see 
\cite{hoang2013complexity,dodwell2015hierarchical,beskos2017multilevel,hoel2016multilevel,jasra2020advanced} 
for examples of MLMC and \cite{jasra2018multi,jasra2021multi} for MIMC.
A notable benefit of MC methods is easy {\em parallelizability}, however typically 
MLMC and MIMC methods for inference are much more synchronous, 
or even serial in the case of MCMC. 
A family of rMLMC methods have recently been introduced for inference 
\cite{jasra2021unbiased,jasra2020unbiased,heng2021unbiased},
which largely recover this lost parallelizability, and deliver i.i.d. samples 
that are unbiased with respect to the limiting infinite resolution target distribution 
{\em in the inference context}. 
In other words, the expectation of the resulting estimators are free 
from any approximation error.
The first instance of rMLMC for inference was \cite{franks2018unbiased},
and the context was different to the above work -- in particular, 
consistent estimators are constructed that are free from discretization bias.

The rest of this paper is focused on these novel parallel rMLMC methods for inference,
which are able to achieve the gold standard of Bayesian posterior inference with 
canonical complexity rate $1/$MSE. In the age of data and increasing parallelism of 
supercomputer architecture, these methods are prime candidates to become a staple, 
if not the defacto standard, for inference in data-centric science and engineering applications.
Section \ref{sec:tech} describes some technical details of the methods, 
Section \ref{sec:example} presents a specific motivating example Bayesian inverse problem and
some compelling numerical results, and 
Section \ref{sec:conclusion} concludes with a call to action and roadmap forward for this exciting research program.

\section{Technical Details of the methodology}\label{sec:tech}

The technical details of the methodology will be sketched in 
this section. The idea is to give an accessible overview and invitation
to this exciting methodology. 
The interested reader can find details in the references cited. 
With respect to the previous section, the notation for $\theta$
will be suppressed -- 
the concerned reader should imagine either everything is conditioned on $\theta$
or it has been absorbed into $u \leftarrow (u,\theta)$.
Subsection \ref{sec:mlmc} sketches the MLMC idea, 
and some of the challenges, strategies for overcoming them, and opportunities
in the context of inference.
Subsection \ref{sec:rmlmc} sketches the rMLMC idea, 
and some of the challenges, strategies for overcoming them, and opportunities
in the context of inference.
Finally subsection \ref{sec:mimc} briefly sketches MIMC.

\subsection{Multilevel Monte Carlo}\label{sec:mlmc}

\begin{figure}
\centering
\subfigure[
Few high fidelity (high-cost) simulations are combined with many at low-fidelity (low cost).]{\includegraphics[width=0.37\textwidth]{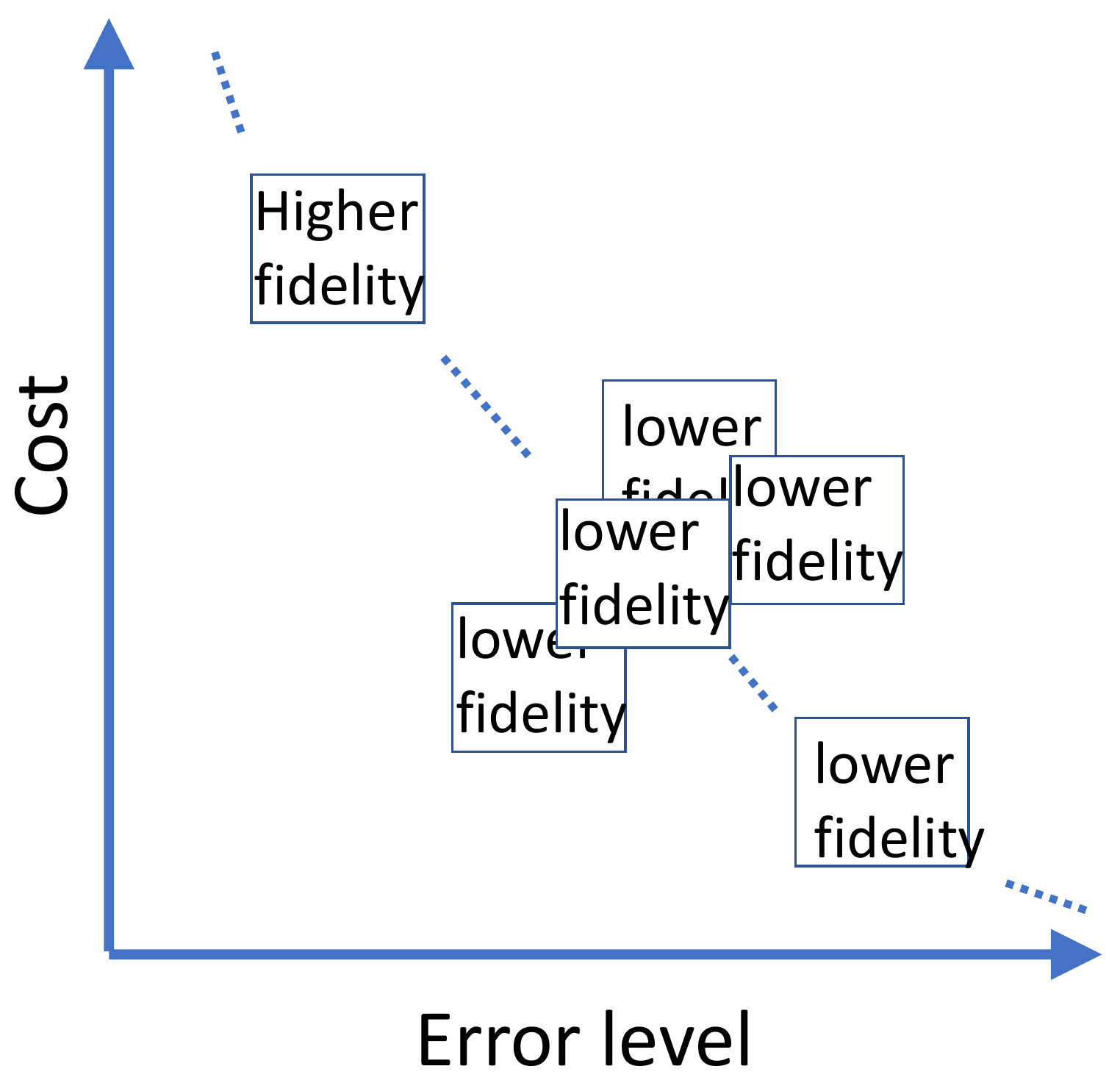}\label{fig:mlmc}}
\hspace{5pt}\subfigure[MSE vs Cost: MLSMC vs SMC for elliptic PDE, illustrating the large gain in efficiency 
(smaller Cost for a given MSE). \cite{beskos2017multilevel}]
{\includegraphics[width=0.57\textwidth]{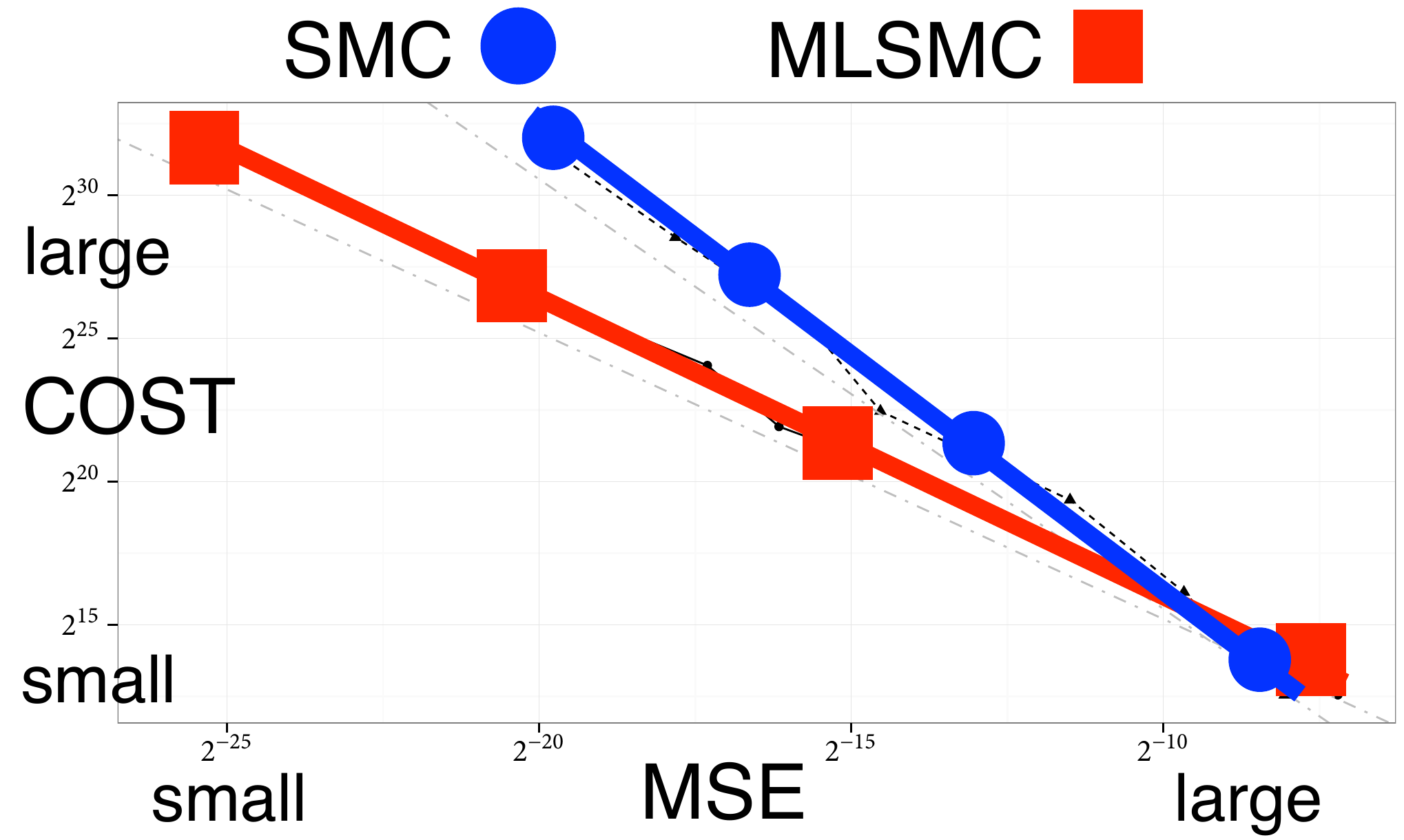}\label{fig:converge}}
\caption{Synopsis of MLMC methods: 
a family of models, 
including coarse-resolution approximation of differential equations, surrogates, etc. (a)
can be combined in the MLMC framework to yield improved complexity cost $\propto 1/$MSE (b).} 
\label{fig:one}
\end{figure}

As mentioned above, for problems requiring approximation, 
MLMC methods are able to achieve a {\em huge speedup}
in comparison to the naive approach of using a single fixed approximation,
and indeed in some cases canonical complexity of cost $\propto1/$MSE. 
These methods leverage a range of successive 
approximations of increasing cost and accuracy.
In a simplified description, most
MLMC theoretical results rely on underlying assumptions of 
\begin{itemize}
\item[(i)] a hierarchy
of targets $\Pi_l$, $l\geq 0$, of increasing cost, such that $\Pi_l \rightarrow \Pi$ as 
$l \rightarrow \infty$ ; 
\item[(ii)] a coupling 
$\Pi^l$ 
s.t. 
$\forall$ 
$A\subset U$, 
$$
\int_{A \times U} \Pi^l (du,du') = \Pi_l(A) \, , \quad {\sf and} \quad
\int_{U \times A} \Pi^l (du,du') = \Pi_{l-1}(A);$$
\item[(iii)] the coupling is such that
\begin{equation}\label{eq:forwardvar}
\int |\varphi(u) - \varphi(u')|^2  \Pi^l (du,du') \leq C h_l^{\beta} \, ,
\end{equation}
and the cost to simulate 
from $\Pi^l$ is proportional to $C h_l^{-\zeta}$, for some $h_l>0$ s.t.  
$h_l \rightarrow 0$ as ${l \rightarrow \infty}$,
and $C, \beta, \zeta >0$ independent of $l$.
\end{itemize}
Now one leverages the telescopic sum 
\begin{equation}\label{eq:telescope}
\Pi(\varphi) = 
\underbrace{\sum_{l=0}^L \Delta_l(\varphi)}_{\sf approximation}  + 
\underbrace{\sum_{l=L+1}^\infty \Delta_l(\varphi)}_{\sf bias} \, ,
\end{equation}
where 
$\Delta_l(\varphi)=\Pi_l (\varphi) - \Pi_{l-1}(\varphi)$, $\Pi_{-1}\equiv 0$,
by approximating the first term, $\Pi_L(\varphi)$, 
using i.i.d. samples from the couplings $\Pi^l$, $l=0,\dots,L$.
The second term is the bias$=\Pi(\varphi) - \Pi_L(\varphi)$. 
This allows one to optimally balance cost with more samples on coarse/cheap levels, 
and a decreasing number of samples as $l$ increases,
to construct a multilevel estimator $\widehat\Pi(\varphi)$ that
achieves a given mean square error (MSE),
$$\bbE ( \widehat\Pi(\varphi) - \Pi(\varphi))^2 = {\sf variance} +{\sf bias}^2 \, ,$$
more efficiently than a single level method. 
A schematic is given in Fig. \ref{fig:mlmc}.

The MLMC estimator is defined as 
\begin{equation}\label{eq:mlest}
\widehat{Y} = \sum_{l=0}^L \frac1{N_l}\sum_{i=1}^{N_l} Y_l^i \, ,
\end{equation}
where $Y_l^i=\varphi(U_l^i) - \varphi(U_{l-1}^i)$ and $(U_l,U_{l-1})^i \sim \Pi^l$ 
for $l\geq 1$, $Y_0^i=\varphi(U_0^i)$, $U_0^i \sim \Pi_0$, 
and $L$ and $\{N_l\}_{l=0}^L$ are chosen to balance the bias and variance.
In particular, $L \propto \log($MSE$)$ and $N_l \propto h_l^{(\beta+\zeta)/2}$.
In the canonical regime where $\beta>\zeta$ one achieves the canonical 
complexity of cost $\propto 1/$MSE. If $\beta \leq \zeta$, there are penalties.
See \cite{giles2015multilevel} for details. 
Note that for the theory above, controlling the bias requires only $\alpha>0$ such that
$$
\left |\int \varphi(u) - \varphi(u')  \Pi^l (du,du') \right| \leq C h_l^{\alpha} \, ,
$$
however it is clear that Jensen's inequality provides $\alpha \geq \beta/2$,
which is suitable for the purposes of this exposition.
There are notable exceptions where one can achieve $\alpha > \beta/2$,
e.g. Euler-Maruyama or Milstein simulation of SDE \cite{giles2008multilevel}, 
and this of course provides tighter results.

Note that the assumptions above can be relaxed substantially if one sacrifices 
a clean theory. In particular, the models $\Pi_l$ need not be defined hierarchically in terms 
of a small parameter $h_l$ corresponding to ``resolution'', as long as $h_l^\beta$
and $h_l^{-\zeta}$ in assumption (iii) above can be replaced with $V_l$ and $C_l$, respectively, 
such that $V_l \rightarrow 0$ as $C_l \rightarrow \infty$ in some fashion.
Indeed in practice one need not ever consider the limit and can work with a finite 
set of models within the same framework, as is advocated in the 
related multifidelity literature (see e.g. \cite{multifidelity}).

\subsubsection{MLMC for inference. }\label{sec:mlmcin}

In the context of inference, it is rare that one can achieve i.i.d. samples 
from couplings $\Pi^l$. As described in Section \ref{sec:intro},
one more often only has access to (unbiased estimates of) the un-normalized target
and must resort to MCMC or SMC.
In the canonical regime $\beta>\zeta$
the theory can proceed in a similar fashion provided one can obtain estimators
$\hat{Y}_l^N$ such that for some $C, \beta >0$ and $q=1,2$ 
\begin{equation}\label{eq:mliass}
\bbE \left [ \hat{Y}_l^N - (\Pi_l(\varphi) - \Pi_{l-1}(\varphi)) \right ]^q \leq C \frac{h_l^{\beta q/2}}{N} \, .
\end{equation}
In the sub-canonical regime, the situation is slightly more complex.

Achieving such estimates with efficient inverse MC methods has been 
the focus of a large body of work. These methods
can be classified according to 3 primary strategies: importance sampling 
\cite{hoang2013complexity,beskos2017multilevel,beskos2018multilevel,moral2017multilevel,jasra2016forward},
coupled algorithms 
\cite{dodwell2015hierarchical,hoel2016multilevel,jasra2017multilevel,gregory2016multilevel,jasmcmc}, 
and approximate couplings \cite{jasra2018bayesian,jasra2018multi,jasra2021multi}. 
See e.g. \cite{jasra2020advanced} for a recent review.
Importance sampling estimators are the simplest, and 
they proceed by expressing the desired increment in terms of 
expectation with respect to one of the levels.
Its applicability is therefore limited to cases where the importance 
weights can be calculated or estimated.
Coupled algorithms attempt to achieve the required rates by 
coupling two single level algorithms targeting the coarse and fine targets, respectively.
These are in some sense the most natural, and in principle the most general,
but it can be deceptively tricky to get them to work correctly.
Approximate coupling is the most straightforward strategy
and can also be quite versatile. In this case, one abandons 
exactness with respect to coarse and fine marginals, and aims
only to achieve well-behaved weights associated to a change of 
measure with respect to an approximate coupling.

\subsection{Randomized Multilevel Monte Carlo}\label{sec:rmlmc}

Randomized MLMC (rMLMC) is defined similarly to \eqref{eq:mlest}
except with a notable difference. 
Define a categorical distribution ${\bf p}=({\bf p}_0,{\bf p}_1,\dots)$ on $\bbZ_+$
and let $L^i \sim {\bf p}$, and $Y_{L^i}^i$ as above.
The single term estimator \cite{rhee2015unbiased} is defined as 
\begin{equation}\label{eq:single}
Z^i = \frac{Y^i_{L^i}}{{\bf p}_{L^i}} \, .
\end{equation}
Notice that, as a result of \eqref{eq:telescope},
 $$\bbE Z^i = \sum_{l=0}^\infty {\bf p}_{l} \bbE  \left (\frac{Y^i_{l}}{{\bf p}_{l}}  \right) = 
 \Pi_0(\varphi) + \sum_{l=1}^\infty \Pi_l(\varphi) - \Pi_{l-1}(\varphi) = \Pi(\varphi) \, ,$$ 
 i.e. this estimator is free from discretization bias.
The corresponding rMLMC estimator is given by
\begin{equation}\label{eq:rmlest}
\widehat{Z} = \frac1{N} \sum_{i=1}^N Z^i = \sum_{l=0}^\infty \frac1{N {\bf p}_l} \sum_{i; L_i=l} Y_l^i \, .
\end{equation}
It is easy to see that $\bbE \#\{i;L_i=l\} = N {\bf p}_l$ and 
$\#\{i;L_i=l\} \rightarrow N {\bf p}_l$ as $N \rightarrow \infty$,
and the optimal choice level of distribution is analogous to level selection above, 
${\bf p}_l \propto N_l$, with $N_l$ as in \eqref{eq:mlest}.
Despite the infinite sum above, this estimator does not incur infinite cost
for finite $N$, because only finitely many summands are non-zero.
Furthermore, ${\bf p}_l \rightarrow 0$, so higher levels are simulated rarely 
and the expected cost is also typically finite.
See \cite{rhee2015unbiased} for further details and other variants.

\subsubsection{rMLMC for inference}\label{sec:rmlmcin}

In the inference context, one typically does not have access to 
{\em unbiased} estimators of $Y_{L^i}$, and rather $\bbE(\hat{Y}_l^N) \neq \Pi_l(\varphi) - \Pi_{l-1}(\varphi)$.
In the finite $L$ case, one can get away with this provided 
\eqref{eq:mliass} holds, however rMLMC methods rely on this property.
In the work \cite{franks2018unbiased},
SMC is used to construct unbiased estimators of increments with respect to 
the {\em un-normalized} target 
(a well-known yet rather remarkable feature of SMC methods \cite{del2004feynman}),
and subsequently a ratio estimator is used for posterior expectations, which are hence 
biased (for finite $N$) but consistent (in the limit $N\rightarrow \infty$) 
with respect to the infinite-resolution ($L = \infty$) target.
Subsequently it has been observed that another inner application of the methodology
presented above in Section \ref{sec:rmlmc} allows one to {\em transform a consistent estimator into
an unbiased estimator} \cite{jasra2020unbiased,jasra2021unbiased}.

In particular, suppose one can couple two estimators 
$\hat{Y}_l^N$ and $\hat{Y}_l^{N'}$,
with $N'>N$, that marginally satisfy \eqref{eq:mliass}, 
and such that the resulting estimator 
satisfies, for $q=1,2$,
\begin{equation}\label{eq:mlidran}
\bbE \left [ \hat{Y}_l^N - \hat{Y}_l^{N'} \right ]^q \leq C \frac{h_l^{\beta q/2}}{N} \, .
\end{equation}
Introduce inner levels $N_k$, $k \geq 1$, such that $N_k \rightarrow \infty$ as 
$k \rightarrow \infty$, and another categorical distribution $\mathsf{p}=(\sf{p}_0,\sf{p}_1,\dots)$ on $\bbZ_+$.
Now let $K^i \sim {\sf p}$, $L^i \sim {\bf p}$ and simulate 
$\hat{Y}_{L^i}^{N_{K^i}}, \hat{Y}_{L^i}^{N_{K^i-1}}$ as above.
The resulting {\em doubly-randomized} single term estimator is given by
\begin{equation}\label{eq:dubsingle}
Z^i = \frac{1}{{\bf p}_{L^i} {\sf p}_{K^i}} \left ( {\hat Y^{N_{K^i}}_{L^i}} - {\hat Y^{N_{K^i-1}}_{L^i}} \right) \, .
\end{equation}
Now, as above,
$$
\bbE \left[ \frac1{{\sf p}_{K^i}} \left ( {\hat Y^{N_{K^i}}_{l}} - {\hat Y^{N_{K^i-1}}_{l}} \right)  \right ]
= \Pi_l(\varphi) - \Pi_{l-1}(\varphi) \, ,
$$
and hence $\bbE Z^i = \Pi(\varphi)$.
Furthermore, the estimators \eqref{eq:dubsingle} can be simulated i.i.d.
In other words, the embarrassingly parallel nature of classical MC estimators
is restored, as well as all the classical results relating to i.i.d. random variables,
such as the central limit theorem.

The work \cite{jasra2020unbiased} leverages such a doubly randomized estimator
for online particle filtering in the framework of \cite{jasra2017multilevel}.
The work \cite{jasra2021unbiased} uses a so-called {\em coupled sum} variant
in the framework of MLSMC samplers \cite{beskos2017multilevel}.
Both of these estimators suffer from the standard limiting MC 
convergence rate with respect to the inner randomization, which is 
sub-canonical. In other words the cost to achieve an estimator
at level $K$ is $\cO(N_K)$ and the error is $\cO(N_K^{-1})$.
As a result, it is not possible to achieve finite variance 
and finite cost, and one must settle for finite variance and 
finite cost {\em with high probability} \cite{rhee2015unbiased}.
In practice, one may truncate the sum at finite $K_{\rm max}$
to ensure finite cost, and accept the resulting bias.

\subsubsection{rMLMCMC}\label{sec:rmlmcin}

An alternative incarnation of the inner randomization can 
be used in the context of MCMC, relying on the unbiased MCMC
introduced in \cite{jacob2020unbiased}, which is based on the approach of \cite{glynn2014exact}.
In \cite{jacob2020unbiased} one couples a pair of MCMCs $(U_n,U'_n)$ targeting the 
same distribution $\Pi$ in such a way that they 
(i) have the same distribution at time $n$,
$U_n \stackrel{\calD}{\sim} U'_{n+1}$, 
(ii) meet in finite time $\bbE(\tau) < \infty$, 
$\tau= \inf \{ n ; U_n = U'_n\}$, and 
(iii) remain identical 
thereafter. 
An unbiased estimator is then obtained via
\begin{eqnarray*}
\widehat{X} &=& \varphi(U_{n^*}) + \sum_{n=n^*+1}^\infty \varphi(U_n) - \varphi(U_n') \\
&=& \varphi(U_{n^*}) + \sum_{n=n^*+1}^\tau \varphi(U_n) - \varphi(U'_n) \, .
\end{eqnarray*}
It is clear that in expectation the sum telescopes, giving the correct expectation 
$\bbE \widehat X = \bbE(\varphi(U_\infty)) = \Pi(\varphi)$.
Such estimators can be simulated i.i.d., which removes the fundamental serial 
roadblock of MCMC, and the finite meeting time ensures 
finite cost. Variations of the approach allow similar efficiency to a single MCMC
for a single CPU implementation, i.e. without leveraging parallelization.
As above, with parallel processors, the sky is the limit.

In order to apply such technology to the present context, 
one couples a pair of coupled chains $(U_{n,l},U_{n,l-1},U_{n,l}',U_{n,l-1}')$
such that 
$$U_{n,l},U_{n,l-1} \stackrel{\calD}{\sim} U_{n+1,l}',U_{n+1,l-1}' \, ,$$
yielding a foursome that is capable of delivering finite-cost unbiased 
estimators of $\Pi_l(\varphi) - \Pi_{l-1}(\varphi)$. Indeed we are also able to 
achieve estimates of the type in \eqref{eq:mliass}, and therefore (for suitable $\beta$)
rMLMC estimators {\em with finite variance and finite cost}. 
Note that only the intra-level pairs need to meet and remain faithful.
Ultimately, the i.i.d. estimators have the following form. 
Simulate $L^i \sim {\bf p}$ as described in Section \ref{sec:rmlmc}, 
and define $Z^i = \widehat{Y}^i_{L^i}/{\bf p}_{L^i}$, where
\begin{eqnarray}\nonumber
\widehat{Y}^i_l &=& \varphi(U_{n^*,l}) - \varphi(U_{n^*,l-1}) \\
\label{eq:ubmcmc1}
&+& 
\sum_{n=n^*+1}^{\tau_l} \varphi(U_{n,l})- \varphi(U_{n,l}') 
- \sum_{n=n^*+1}^{\tau_{l-1}} 
\left (\varphi(U_{n,l-1}) - \varphi(U'_{n,l-1}) \right) \, , 
\end{eqnarray}
with $\tau_ \ell = \inf \{ n ; U_{n,\ell} = U'_{n,\ell}\}$, for $\ell=l,l-1$.
The final estimator is 
\begin{equation}
\widehat{Z} = \frac1N \sum_{i=1}^N Z^i \, .
\label{eq:ubmcmc2}
\end{equation}

\subsection{Multi-index Monte Carlo}\label{sec:mimc}

Recently, 
the hierarchical telescopic sum identity that MLMC is based
upon has been viewed through the lense of sparse grids, 
for the case in which there are multiple continuous spatial, 
temporal, and/or parametric dimensions of approximation \cite{haji2016multi}.  
In other words, 
there is a hierarchy
of targets $\Pi_\alpha$, where $\alpha$ is a multi-index, 
such that $\Pi_\alpha \rightarrow \Pi$ as $|\alpha| \rightarrow \infty$.
Under a more complex set of assumptions,
one can appeal instead to the 
identity 
$$\Pi(\varphi) = \sum_{\alpha \in \cI} \Delta_\alpha (\varphi) + 
{\sum_{\alpha \notin \cI} \Delta_\alpha (\varphi)}
\, , \quad \cI \subset \mathbb Z_+^d \, ,  $$
where $d-$fold multi-increments $\Delta_\alpha$ are used instead, i.e.
letting $e_j \in \bbR^d$ denote the $j^{\rm th}$ standard basis vector and
$\delta_{j}\Pi_\alpha := \Pi_\alpha - \Pi_{\alpha-e_j}$, then
$\Delta_\alpha := \delta_d \circ \cdots \circ \delta_1 \Pi_\alpha$ 
(for any multi-index $\alpha'$ with $\alpha'_i <0$ for some $i=1,\dots,d$, $\Pi_{\alpha'}:=0$).
The first term 
is approximated again using coupled samples and the second is the bias.
Under suitable regularity conditions, this MIMC 
method {\em yields further huge speedup} 
to obtain a given level of error \cite{giles2015multilevel,haji2016multi}.
Some preliminary work in this direction has been done recently 
\cite{jasra2018multi,jasra2021multi}.
Forward randomized MIMC (rMIMC) has recently been done as well \cite{crisan2018unbiased}.


\section{Motivating example}\label{sec:example}

\subsection{Example of Problem}\label{ssec:example}

The following particular problem is presented as an example.
This example is prototypical of a variety of inverse problems involving
physical systems in which noisy/partial observations are made of the solution of an elliptic PDE
and one would like to infer the diffusion coefficient. 
For example, the solution to the PDE $v$ could represent pressure of a patch of land, 
subject to some forcing $f$ (sources/sinks), 
and the diffusion coefficient $\hat{u}(u)$ then corresponds to the subsurface permeability \cite{tarantola,stuart},
a highly desirable quantity of interest in the context of oil recovery. 
Let $D\subset\mathbb{R}^d$ with $\partial D\in C^1$ convex
and $f\in L^2(D)$. 
Consider the following PDE on $D$:

\begin{align}
-\nabla \cdot (\hat{u}(u)\nabla v)  &=f,\quad \textrm{ on } D,       \\
v&= 0, \quad \textrm{ on } \partial D \, ,
\label{eq:pde}
\end{align}
{where the diffusion coefficient has the form
\begin{equation}
\hat{u}(x; u) = 
\label{eq:exuni}
 \bar{u} + \sum_{j=1}^J u_j\sigma_j\phi_j(x) \, , 
\end{equation}
Define $u=\{u_j\}_{j=1}^J$,
and the state space will be $\mathsf{X}=\prod_{j=1}^J[-1,1]$. 
Let $v(\cdot;u)$ denote the weak solution of $(1)$ for parameter value $u$. 
The prior is given by $u_j \sim U[-1,1]$  
(the uniform distribution on $[-1,1]$) i.i.d. for $j=1,\dots, J$. 
It will be assumed that $\phi_j \in C(D)$, $\|\phi_j\|_\infty \leq 1$, 
and there is a $u_*>0$ such that 
$\bar{u} > \sum_{j=1}^J\sigma_j + u_*$. 
Note that under the given assumptions, 
$\hat u(u) > u_*$ uniformly in $u$. 
Hence there is a well-defined (weak) solution $v(\cdot ;u)$
that is bounded in $L^\infty(D)$ and $L^2(D)$ uniformly in $u$, 
and its gradient is also bounded 
in $L^2(D)$ uniformly in $u$ \cite{ciarlet,dashti}.
\\ \indent
Define the following vector-valued function 
\begin{equation}\label{eq:gee}
G(u) = [\langle g_1, v(\cdot;u)\rangle,\dots,\langle g_m, v(\cdot;u)\rangle ]^{\intercal},
\end{equation}
where $g_i \in L^2(D)$ 
for $i=1,\dots,m$. 
We note that pointwise evaluation is also permissible since $u \in L^\infty(D)$, 
i.e. $g_i$ can be Dirac delta functions, 
however for simplicity we restrict the presentation to $L^2(D)$. 
It is assumed that the data take the form 
\begin{equation}\label{eq:obs}
y = G(u) + \xi,\quad \xi \sim N(0,\theta^{-1}\cdot\bm{I}_m),\quad \xi 
\perp u \, ,
\end{equation}
where 
$\perp$ 
denotes independence. 
The unnormalized density 
$\gamma_{\theta}: \mathsf X \rightarrow \bbR_+$
of $u$ for fixed $\theta>0$ is given by 
\begin{equation}\label{eq:unno}
\gamma_{\theta}(u) = 
\theta^{m/2}\exp\Big(-\frac{\theta}{2}\|G(u) - y\|^2\Big) 
\, . 
\end{equation}
The normalized density is 
$$
\eta_{\theta}(u) = \frac{\gamma_{\theta}(u)} {I_{\theta}} \, ,
$$
where $I_{\theta} = {\int_{\mathsf X}\gamma_{\theta}(u)du}$, 
and the quantity of interest is defined for $u \in \mathsf{X}$ as 
\begin{equation}\label{eq:phi}
\varphi_{\theta}(u) := \nabla_{\theta}\log\Big(\gamma_{\theta}(u)\Big) =
 \frac{m}{2\theta} - \frac{1}{2}\|G(u) - y\|^2 \, . 
\end{equation}

To motivation this particular objective function, 
notice that $\gamma_\theta$ is chosen such 
that the marginal likelihood, or ``evidence'' for $\theta$, is given 
by $p(y|\theta) = I_\theta$.
Therefore the MLE ($\lambda=0$) or MAP are given as minimizers of 
$-\log I_\theta + \lambda R(\theta)$, where $R(\theta)= -\log p(\theta)$.
Assuming $R(\theta)$ is known in closed form and differentiable, 
then a gradient descent method requires 
\begin{equation}\label{eq:gradest}
\nabla_\theta \log I_\theta =  \frac1{I_\theta} \int_{\mathsf X} 
\nabla_{\theta}\gamma_{\theta}(u) du =
\frac1{I_\theta} \int_{\mathsf X} 
\underbrace{\nabla_{\theta}\log\Big(\gamma_{\theta}(u)\Big)}_{\varphi_{\theta}(u)} 
\gamma_{\theta}(u) du = \eta_\theta (\varphi_{\theta}(u)) \, .
\end{equation}
Stochastic gradient descent requires only an unbiased estimator
of $\eta_\theta (\varphi_{\theta}(u))$ \cite{kushner2003stochastic}, 
which the presented rMLMC method delivers.
}


\subsubsection{{Numerical approximation}}\label{sssec:example}

{The finite element method (FEM) is utilized for solution of \eqref{eq:pde}
with piecewise multi-linear nodal basis functions.
Let $d=1$ and $D=[0,1]$ for simplicity.
Note the approach is easily generalized to $d \geq 1$
using products of such piecewise linear functions described below
following standard FEM literature \cite{brenner}.}
The PDE problem at resolution level $l$ is solved using FEM
with piecewise linear shape functions on a uniform mesh of width {$h_l=2^{-l}$, for $l\geq0$}. 
Thus, on the $l$th level the finite-element basis functions are $\{\psi_i^l\}_{i=1}^{2^l-1}$ defined as (for $x_i = i\cdot 2^{-l}$):
$$
\psi_i^l(x) =  \left\{\begin{array}{ll}
(1/h_l)[x-(x_i-h_l)]
& \textrm{if}~x\in[x_i-h_l,x_i], \\
(1/h_l)[x_i+h_l-x] & \textrm{if}~x\in[x_i,x_i+h_l] \, .
\end{array}\right.
$$
To solve the PDE, $v^l(x)=\sum_{i=1}^{2^l-1}v_i^l\psi_i^l(x)$ is plugged into (1), 
and projected onto each basis element:
$$
-\Big\langle \nabla\cdot\Big(\hat{u}\nabla\sum_{i=1}^{2^l-1}v_i^l \psi_i^l \Big),\psi_j^l \Big\rangle = \langle f, \psi_j^l \rangle, 
$$
resulting in the following linear system:
$$
\bm{A}^l(u)\bm{v}^l = \bm{f}^l,
$$
where we introduce the matrix $\bm{A}^l(u)$ with entries $A_{ij}^l(u) = \langle \hat{u}\nabla\psi_i^l,\nabla\psi_j^l \rangle$, 
and vectors $\bm{v}^l, \bm{f}^l$ with entries $v_i^l=\langle v, \psi_i^l\rangle$ and $f_i^l=\langle f, \psi_i^l\rangle$, respectively.

Define $G^l(u) = [\langle g_1, v^l(\cdot;u)\rangle,\dots,\langle g_m, v^l(\cdot;u)\rangle]^{\intercal}$.
Denote the corresponding approximated un-normalized density by 
\begin{equation}\label{eq:unnol}
\gamma_{\theta}^l(u) = \theta^{m/2}\exp\Big\{-\frac{\theta}{2}\|G^l(u) - y\|^2\Big\} \, ,
\end{equation}
and the approximated normalized density by
$\eta_{\theta}^l(u) = {\gamma_{\theta}^l(u)} / {I_\theta^l}$, 
where $I_\theta^l = {\int_{\mathsf{X}}\gamma_{\theta}^l(u)du}$.
Furthermore, define 
\begin{equation}\label{eq:phil}
\varphi^l_\theta(u) := \nabla_{\theta}\log\Big(\gamma_{\theta}^l(u)\Big) 
= \frac{m}{2\theta} - \frac{1}{2}\|G^l(u) - y\|^2 \, .
\end{equation}

It is well-known that under the stated assumptions 
$v^l(u)$ converges to $v(u)$ as $l \rightarrow \infty$ in $L^2(D)$ (as does its gradient), 
uniformly in $u$ \cite{brenner,ciarlet}, with the rate $h_l^{\beta/2}$, $\beta=4$.
In a forward UQ context, 
this immediately provides \eqref{eq:forwardvar} for Lipschitz functions of $v$, with $\beta=4$.
Furthermore, continuity ensures $\gamma_{\theta}^l(u)$ 
converges to $\gamma_{\theta}(u)$ and  
$\varphi^l_\theta(u)$ converges to $\varphi_\theta(u)$ uniformly in $u$ as well. 
See also \cite{beskos,beskos2018multilevel} 
for further details. 
This allows one to achieve estimates of the type \eqref{eq:mliass} in the inference context.

\subsection{Numerical results}\label{sec:numerics}

This section is for illustration purposes and 
reproduces results from \cite{heng2021unbiased}, 
specifically Section 4.1.2 and Figure 5.
The problem specified in the previous section is considered with
forcing $f(x)=100x$. 
The prior specification of $u=(u_1,u_2)$ is taken as $J=2$,
$\bar u=0.15$, $\sigma_1=1/10$, $\sigma_2=1/40$, 
$\phi_1(t)=\sin(\pi x)$ and $\phi_2(t)=\cos(2\pi x)$. 
For this particular setting, the solution $v$ is continuous and 
hence point-wise observations are well-defined. 
The observation function $G(x)$ in \eqref{eq:gee} 
is chosen as $g_i(v(u))=v(0.01+0.02(i-1);u)$ for $i\in\{1,\ldots,m\}$ with $m=50$. 
The FEM scheme in Section \ref{sssec:example} is employed with mesh width of 
$l \leftarrow l+l_0$, where $l_0=3$. 
Using a discretization level of $l=10$ to approximate $G(x)$ with $G_l(x)$, 
$x=(0.6,-0.4)$ and $\theta=1$,  
observations $y\in\mathbb{R}^m$ are simulated from \eqref{eq:obs}. 

\begin{figure}[!htbp]
	\centering\includegraphics[width=1\textwidth]{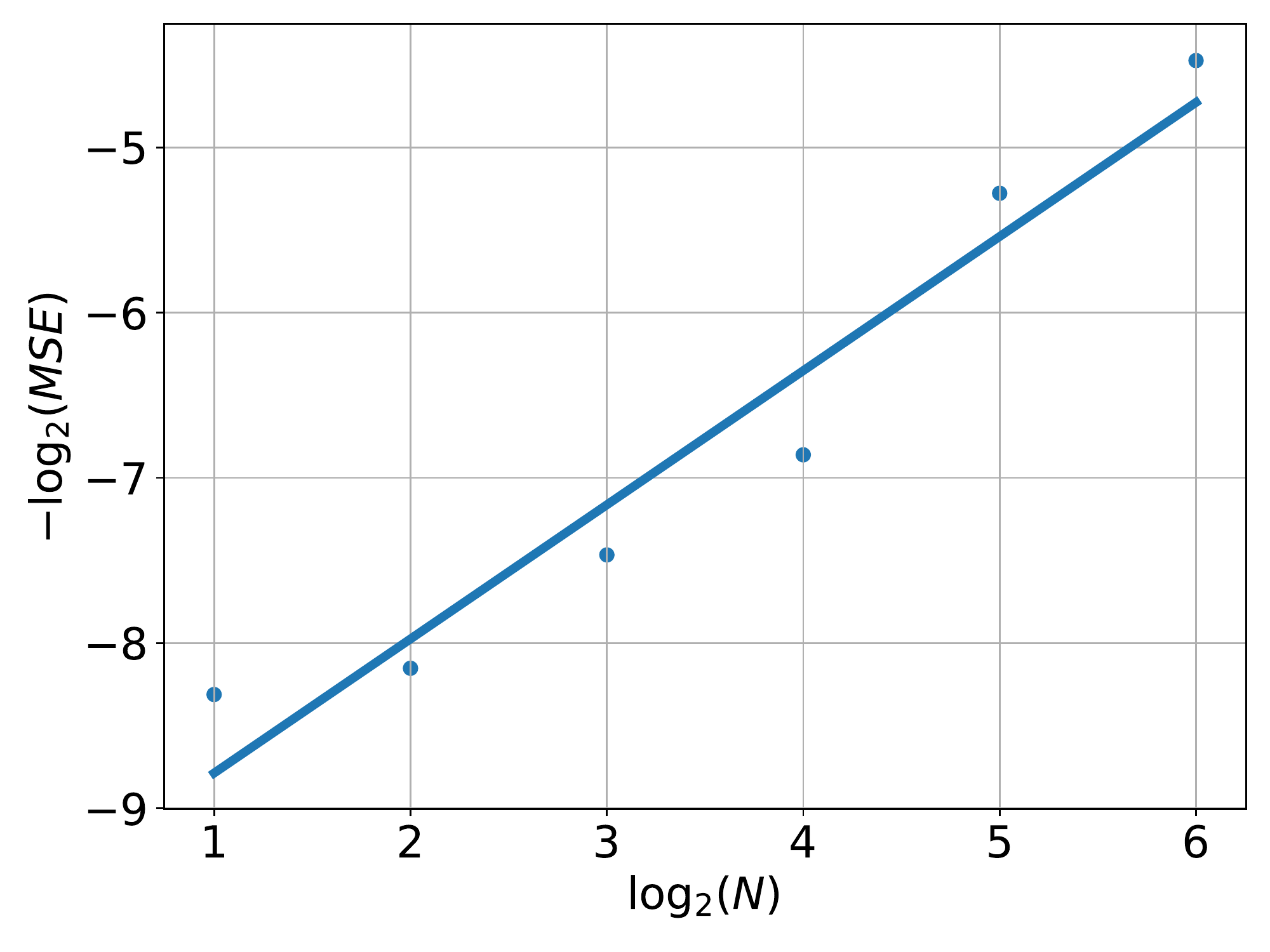}
	\includegraphics[width=1\textwidth]{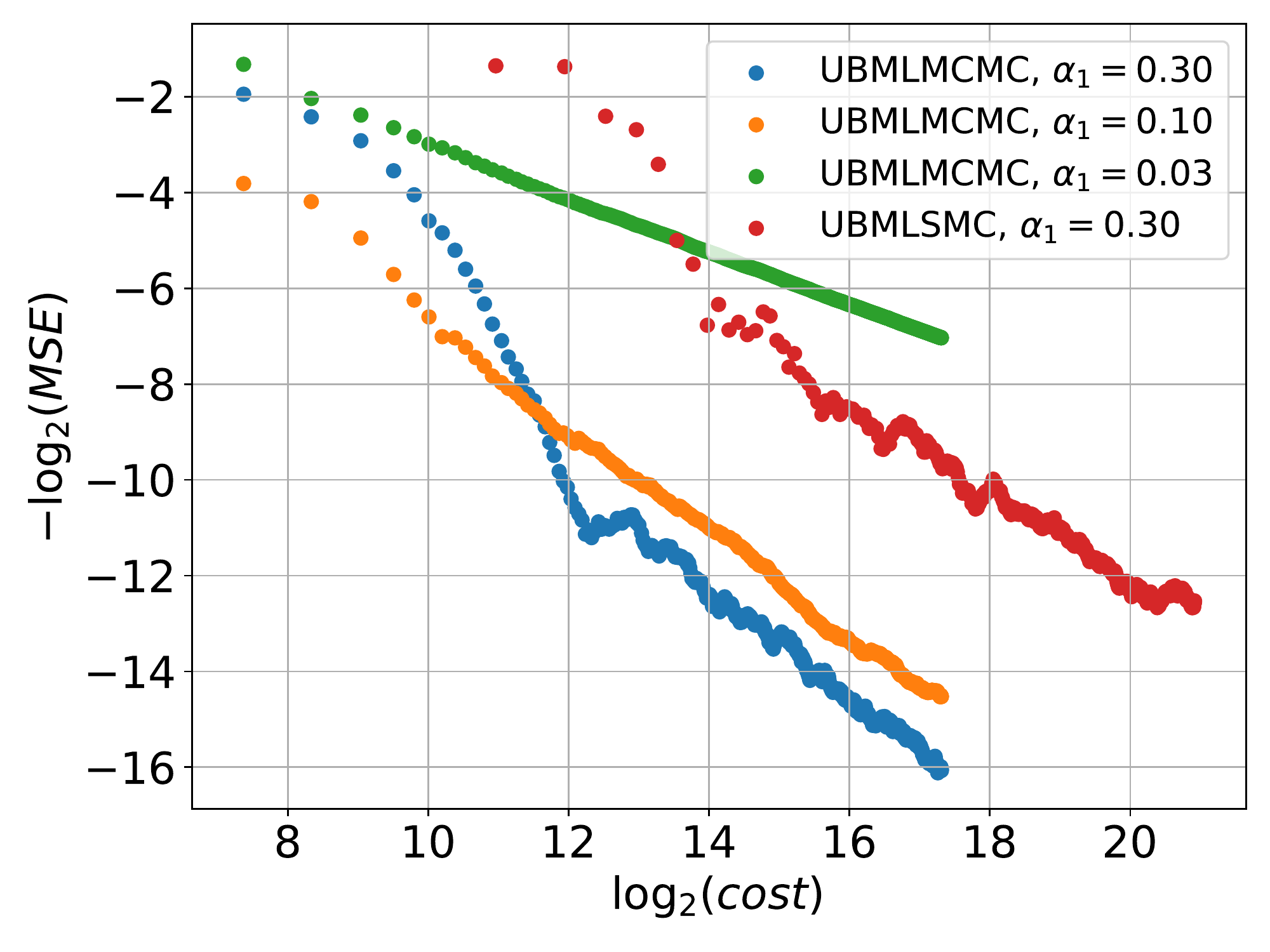}
	\caption{Elliptic Bayesian inverse problem of Section \ref{sec:numerics} 
	Left: accuracy (minus MSE) 
	against number of single term samples $N$. 
	The samples were simulated in serial on a laptop, but can all be simulated in parallel.
	Right: convergence of stochastic gradient iterates $\theta^{(n)}$ 
	to the maximum a posteriori probability estimator $\theta_{\rm MAP}$.   
	The learning rates considered here are 
	$\alpha_n=\alpha_1/n$. 
	The red curve corresponds to 
	the unbiased 
	MLSMC algorithm of \cite{jasra2021unbiased} for comparison. }	
	\label{fig:nl_mse}
\end{figure}

The estimators $\widehat{Y}^i_{L^i}$ are computed using a reflection maximal coupling of 
pCN kernels, as described in \cite{heng2021unbiased}. 
The left panel of Figure \ref{fig:nl_mse} illustrates that averaging single term estimators 
\eqref{eq:ubmcmc1} as in \eqref{eq:ubmcmc2}
yields a consistent estimator that converges at the canonical Monte Carlo rate of $1/$MSE. 

Consider now inference for $\theta$ in the Bayesian framework, 
under a prior $p(\theta)$ specified as a standard Gaussian prior on $\log \theta$. 
A stochastic gradient ascent algorithm is initialized at $\theta^{(0)}=0.1$ to compute 
the maximum a posteriori probability (MAP) estimator 
$\theta_{\rm MAP}\in \arg\max p(\theta)I_\theta$, 
simulated by subtracting $\nabla_\theta R(\theta)$ from 
the estimator of \eqref{eq:gradest} given by $Z^i$ defined above and in \eqref{eq:ubmcmc1}. 
The right panel of Figure \ref{fig:nl_mse} displays convergence of the stochastic iterates 
to $\theta_{\rm MAP}$. 
An estimator following \cite{jasra2021unbiased}, 
of the type in \eqref{eq:dubsingle},
is also shown here,
using the algorithm in \cite{beskos} instead of coupled MCMC. 
The plot shows some gains over \cite{jasra2021unbiased} 
when the same learning rates are employed.

\subsubsection{Parallel implementation. }\label{ssec:parallel}

An example is now presented to illustrate the 
parallel improvement of these methods 
on multiple cores. 
These results are borrowed from \cite{jasra2020unbiased} for (online)
filtering of partially observed diffusions.
In particular, an estimator of the form \eqref{eq:dubsingle} is constructed, 
in which each $\hat{Y}_{L^i}^{N_{K^i}}$ is a coupled particle filter increment estimator
at resolution $L^i$ and with $K^i$ particles, for $i=1,\dots,N$, 
and  these estimators are then averaged as in \eqref{eq:ubmcmc2}.
The parallel performance 
is assessed with up to $1000 (\leq N)$ MPI cores 
on the KAUST supercomputer Shaheen. 
A Python notebook that implements the unbiased estimator both on a single core and multiple cores can be found in the following Github link: 
\verb|https://github.com/fangyuan-ksgk/Unbiased-Particle-Filter-HPC-|.

To demonstrate the parallel scaling power, various numbers of processors 
$M \in \{1,5,10,20,50,100,500,1000\}$ are used, with $N=10^3 M$. 
The serial computation time to obtain the estimator on a single core is recorded, 
as well as the parallel computation time on $M$ cores. 
The parallel speedup is defined as the ratio of cost for serial implementation and the cost for parallel implementation, and the parallel efficiency is given by the ratio of parallel speedup and the number of parallel cores $M$. 

\begin{figure}[!htbp]
{\includegraphics[width=\textwidth,height=0.7\textwidth]{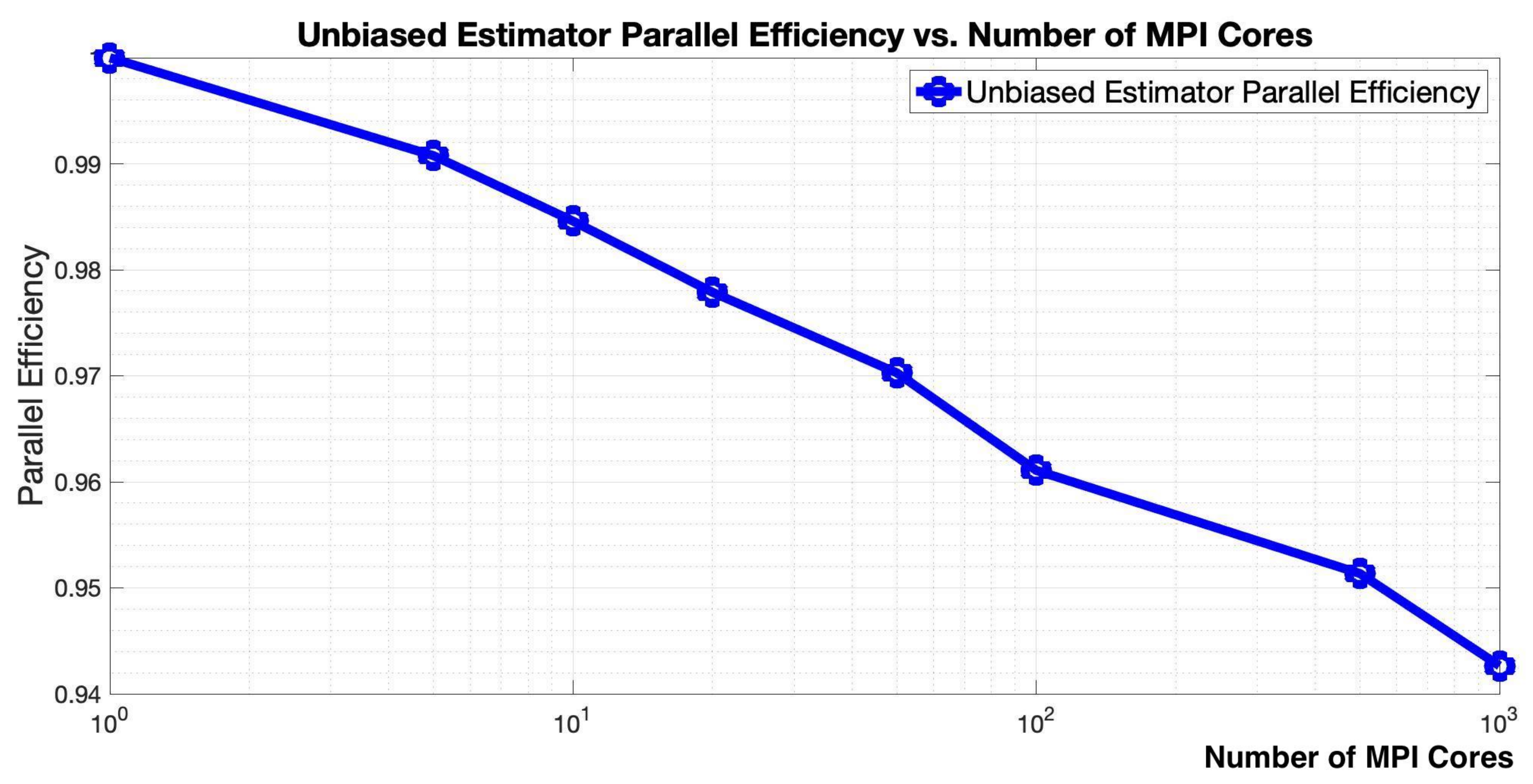}}
{\includegraphics[width=\textwidth,height=\textwidth]{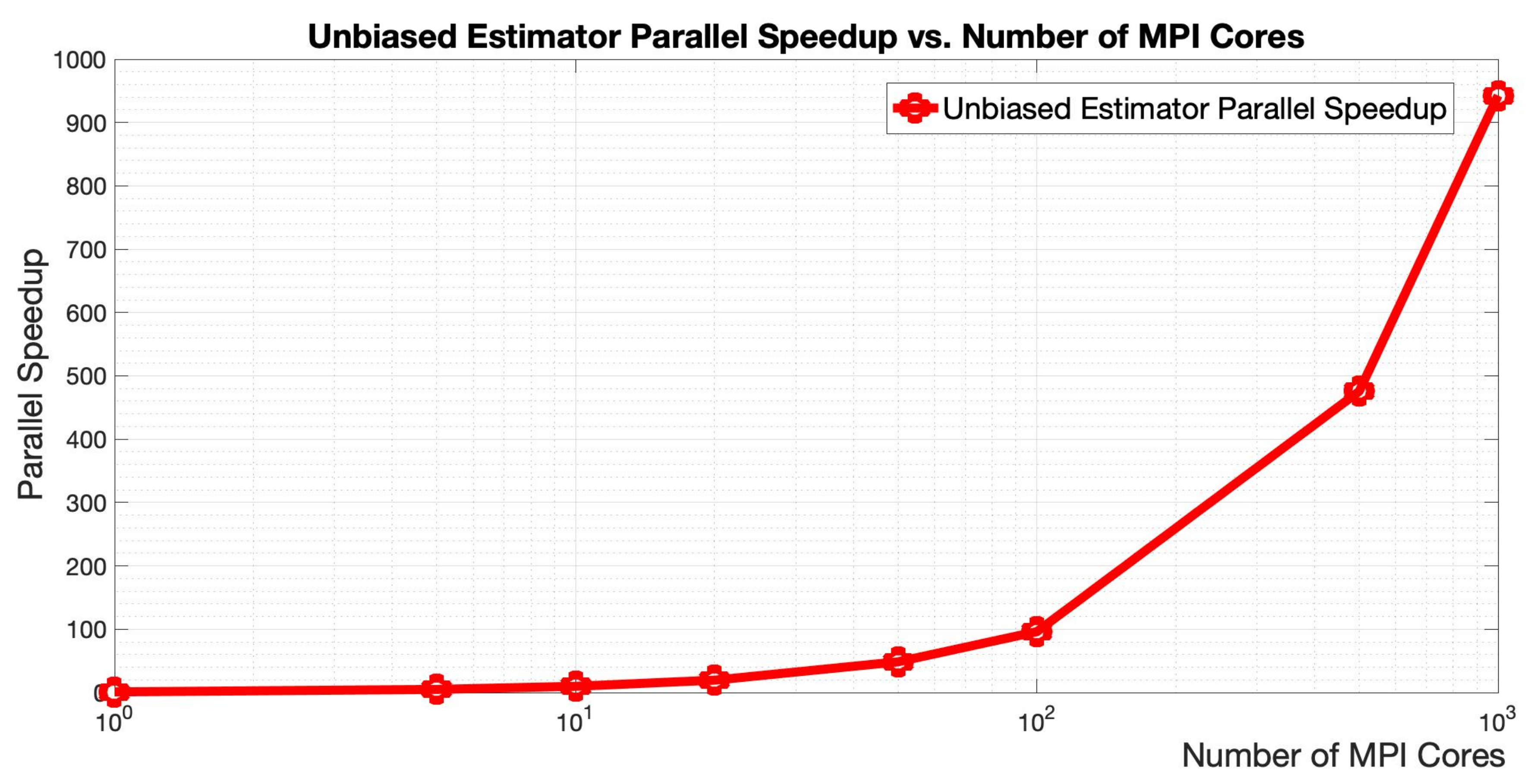}}
\caption{Parallel Speedup and Parallel Efficiency against number of MPI cores 
for the unbiased particle filter from \cite{jasra2020unbiased}.}\label{fig:par}	
\end{figure}

The results are shown in Figure \ref{fig:par}, which shows almost perfect strong scaling 
for up to $1000$ MPI cores, for this level of accuracy. 
It is important to note that there will be a limitation to the speedup possible, 
depending upon the accuracy level. 
In particular, the total simulation time is limited by the single most expensive sample required.
Therefore, it will not be possible to achieve MSE$\propto \varepsilon^2$ 
in $\cO(1)$ time, even with arbitrarily many cores.

\section{Conclusion and path forward}\label{sec:conclusion}

This position paper advocates for the widespread adoption of 
Bayesian methods for performing inference, especially in the context of complex 
science and engineering applications, where 
high-stakes decisions require robustness, generalizability, and interpretability. 
Such methods are rapidly gaining momentum in science and engineering 
applications, following an explosive interest in UQ, in concert with the data deluge and 
emerging fourth paradigm of data-centric science and engineering.
Meanwhile, in the field of machine learning and AI 
the value of Bayesian methods 
has been recognized already 
for several decades. 
There it is widely accepted that the Bayesian posterior is the gold standard, 
but the community has largely converged on variational approximations
or even point estimators as surrogates, due to complexity limitations.

Here a family of embarrassingly parallel rMLMC simulation methods 
are summarized. 
The methods are designed for performing exact Bayesian inference 
in the context where only approximate models are available, 
which includes a wide range of problems in physics, biology, finance, machine learning, 
and spatial statistics. Canonical complexity is achieved.
Important priorities going forward are: 
(i) continued development of novel instances of 
this powerful class of algorithms, 
(ii) adaptation to specific large scale application contexts across 
science, engineering, and AI, and 
(iii) automation of the methods and the
design of usable software to enable deployment 
on a large scale and across applications in science, engineering, and AI, 
ideally by practitioners and without requiring an expert. 

\vspace{10pt}

\noindent {\bf Acknowledgements. } 
KJHL and AT were supported by The Alan Turing Institute 
under the EPSRC grant EP/N510129/1.
AJ and FY acknowledge KAUST baseline support.


%
%
\bibliographystyle{plain}
\bibliography{refs}

%
%
%
%
%
%
%
%
\end{document}